\documentclass[prb, twocolumn, superscriptaddress, longbibliography]{revtex4-2}

\usepackage{graphicx,epsfig} 
\usepackage{amsmath}
\usepackage{amsfonts}
\usepackage{cancel}
\usepackage[colorlinks=true, allcolors=blue]{hyperref}
 \usepackage{soul}
 \usepackage{xcolor}
 
\usepackage{float}
\usepackage[section]{placeins}

\newcommand{\ve}[1]{\boldsymbol{#1}}
\renewcommand{\vec}[1]{\boldsymbol{#1}}

\begin{document}

\title{Finite temperature fermion Monte Carlo simulations of frustrated spin-Peierls systems}

\author{Jo\~ao C. In\'acio}%
\affiliation{\mbox{Institut f\"ur Theoretische Physik und Astrophysik, Universit\"at W\"urzburg, 97074 W\"urzburg, Germany}}
\email{joao.carvalho-inacio@uni-wuerzburg.de}

\author{Jeroen van den Brink}
\affiliation{\mbox{Institute for Theoretical Solid State Physics, IFW Dresden, 01069 Dresden, Germany}}
\affiliation{\mbox{W\"urzburg-Dresden Cluster of Excellence ct.qmat, Germany}}

\author{Fakher F. Assaad}
\affiliation{\mbox{Institut f\"ur Theoretische Physik und Astrophysik, Universit\"at W\"urzburg, 97074 W\"urzburg, Germany}}
\affiliation{\mbox{W\"urzburg-Dresden Cluster of Excellence ct.qmat, Germany}}

\author{Toshihiro Sato}%
\affiliation{\mbox{Institute for Theoretical Solid State Physics, IFW Dresden, 01069 Dresden, Germany}}
\affiliation{\mbox{W\"urzburg-Dresden Cluster of Excellence ct.qmat, Germany}}
		
\date{\today}

\begin{abstract}
	The  Abrikosov fermion representation of the spin-1/2 degree of freedom allows for auxiliary-field quantum Monte Carlo simulations of frustrated spin systems. This approach provides a  manifold of equivalent actions over which the negative sign problem can be optimised. As a result, we can reach temperature scales   well below  the magnetic scale. Here, we show  how to generalise this algorithm to  spin-Peierls systems. In contrast to exact diagonalisation approaches, Monte Carlo methods are not Hilbert space bound such that the computational effort per sweep remains invariant when adding phonons. However, the computational  effort required to generate independent configurations increases in the presence of phonons. We also show that, for the specific case of the Kitaev-Heisenberg model, the inclusion of phonons does not render the negative sign problem more severe. This new algorithm hence allows us to investigate the interplay between phonon degrees of freedom and magnetic frustration. We present results for frustrated and non-frustrated spin systems.
\end{abstract}

\maketitle

\section{Introduction}

Quantum spin liquid states, unlike traditional magnetically ordered states with spin wave modes, are defined by fractionalisation  and concomitant emergence of a deconfined gauge theory \cite{Balents10}. This distinction becomes even more compelling when considering the interaction with dynamical phonons, which can alter phonon excitation behaviours depending on the magnetic excitations within the spin sector. For instance, in the simplest mean-field description of  a  spin liquid in terms of spinons, phonons will exhibit Landau damping when interacting with the two-spinon continuum \cite{Mahan90}. Our study draws significant inspiration from  Raman spectroscopy findings on the Kitaev spin liquid candidate $\alpha$-RuCl$_3$ \cite{Sandilands15a}. These findings indicated a continuum of fractionalised excitations coupled to   (optical) 
phonons, setting the stage for our investigation. Our aim here is to employ exact quantum Monte Carlo (QMC) simulations to uncover potential correlations between phonons and spins in frustrated spin systems. We will specifically focus on optical phonons interacting with exchange interactions in a manner akin to the spin-Peierls fashion.  While we will  concentrate on the Kitaev-Heisenberg Hamiltonian, generalisations to realistic magnetic models of $\alpha$-RuCl$_3$  are  straightforward. 

The spin-1/2 degree of freedom, represented using an Abrikosov fermion supplemented by a single occupancy constraint, provides a path for simulating spin systems through the auxiliary-field QMC (AFQMC) simulations for fermions \cite{Blankenbecler81, White89,Assaad08_rev,ALF_v2}. 
Although this fermionic approach to simulating spin models might appear somewhat contrived, it raises the question of whether it is competitive and a worthy route to pursue.
While world-line based QMC methods, such as the Stochastic Series Expansion (SSE) algorithm \cite{Sandvik02,Assaad08_rev}, are traditionally preferred for simulating negative-sign-free spin and spin-Peierls models \cite{Weber17,Weber21a}, our fermionic QMC strategy has potential for tackling frustrated spin systems. Indeed, our recent advancements in the AFQMC method for aforementioned Kitaev-type models \cite{SatoT21}, address the challenging negative sign problem encountered in standard world-line based QMC implementations. By using this auxiliary-field approach, one can reach temperature scales down  to \(40\text{K}\) for realistic models of  $\alpha$-RuCl$_3$ \cite{SatoT23a}  and thereby confront Monte Carlo results to experimental data. 

The Debye temperature of $\alpha$-RuCl$_3$ is set at 200K such that lattice fluctuations can play an important role. The  aim of this paper is to show that one can generalise the method of  Refs.~\cite{SatoT21,SatoT23a} to  include phonons. Including phonons in Hilbert space-based methods such as exact diagonalisation or density matrix renormalisation group (DMRG) is hard since one  has to confront the infinite size of the  Hilbert space \cite{Stolpp21}. The key point of the  Monte Carlo method is that it is not limited by the size of the Hilbert space. In the absence of fat tails \cite{Hao16,Ulybyshev21_1}, the variance of the result is inversely proportional to the square root of the number of independent  samples one can produce. In the adiabatic limit, it is often hard to produce independent samples, and many approaches have been used. This includes local updates,  global updates \cite{Goetz21}, or integrating out   phonon degrees of freedom at the expense of retarded interactions \cite{Assaad07, Weber17, Weber21a, Karakuzu18, Goetz23}. Here, we will adopt a  local update. The key result of the paper is  that including phonons, in the physically relevant adiabatic limit, does not  increase the negative sign problem. Hence, our AFQMC approach sets a new standard for addressing spin-phonon interactions in frustrated spin systems.

The paper is organised as follows. In Sec. \ref{sec:spin_seierls_AFQMC}, we define the model and the auxiliary-field formulation of the partition function, which forms the basis for the numerical method. In Sec. \ref{sec:results}, we present the results, starting with the sign problem for the frustrated spin system and then a discussion of the physics, focusing on the temperature dependence of the uniform susceptibility. Finally, in Sec. \ref{sec:conclusion}, we conclude our study.

\section{AFQMC formulation of the generic spin-Peierls Hamiltonian}
\label{sec:spin_seierls_AFQMC}

Our starting point is the following spin-Peierls Hamiltonian with  Einstein phonons:
\begin{eqnarray} \label{eq:spin_peierls_hamiltonian}
	\hat{H}  &=& \sum_{b=\langle \ve{i}, \ve{j} \rangle} \left(1 + g \hat{Q}_{b} \right) \left(J_{b}  \vec{\hat{S}}_{\ve{i}}  \cdot \vec{\hat{S}}_{\ve{j}}+\sum_{\alpha, \beta}\Gamma_{b}^{\alpha,\beta} \hat{S}_{\ve{i}}^{\alpha}  \hat{S}_{\ve{j}}^{\beta} \right)  \nonumber \\
	&&+\sum_{b=\langle \ve{i}, \ve{j} \rangle} \left( \frac{\hat{P}_{b}^2}{2 m} + \frac{k}{2} \hat{Q}_{b}^2 \right)
\end{eqnarray}
with \(\alpha, \beta = x,y,z\) and 
\begin{eqnarray} \label{Eq:SN-cr}
		[\hat{S}_{\ve{i}}^{\alpha}, \hat{S}_{\ve{j}}^{\beta}]=i \epsilon_{\alpha,\beta,\gamma}\delta_{\ve{i},\ve{j}}\hat{S}_{\ve{i}}^{\gamma}~~,~~
	[\hat{P}_{b}, \hat{Q}_{b'}]=\frac{\delta_{b,b'}}{i}.
\end{eqnarray}
Here the spin-1/2 operators $\vec{\hat{S}}_{\ve{i}}$ reside on a lattice with sites labeled by $\ve{i},\ve{j}$, with the sum running over nearest-neighbor bonds $b=\langle \ve{i},\ve{j} \rangle$. The phonons with the momenta $\hat{P}_{b}$ and displacements $\hat{Q}_b$  reside on each bond.  A  canonical 
transformation \(\hat{Q}_{b} \to \hat{Q}_b/g\)  and \( \hat{P}_{b} \to  g \hat{P}_b\)  allows us to  set \(g\) to unity by redefining the mass $m$ and spring constant $k$.  With this  choice,   the spin-phonon coupling  $\lambda$,  and 
the phonon  frequency $\omega_0$ read
\begin{eqnarray}
	\lambda = \frac{1}{2k}~~,~~\quad \omega_0 = \sqrt{\frac{k}{m}},
\end{eqnarray}
The spin-phonon interaction is mediated via the $ \Gamma_{b}^{\alpha,\beta}$ and $J_{b} $ exchange couplings. While $ \Gamma_{b}^{\alpha,\beta}$  defines the potentially frustrated spin model, $J_{b} $ accounts for non-frustrating exchange couplings. The special case where \(J_b = J\) and \(\Gamma_b^{\alpha,\beta} = 0\) is the Heisenberg model. Meanwhile, setting \(J_b = J = A \cos(\phi)\) and \(\Gamma_b^{\alpha,\beta} = 2 K \delta_{\alpha, \beta} \delta_{\beta, b} = 2 A \sin(\phi)\) defines the Kitaev-Heisenberg model, with \(A = \sqrt{J^2 + K^2}\). Note that the latter definition is only valid in lattices with a coordination number of \(3\). For the remainder of this work, the Kitaev-Heisenberg model is defined on the honeycomb lattice.

Our AFQMC formulation \cite{SatoT21} is based on the Abrikosov fermion representation of the spin operator 
\begin{eqnarray} \label{Eq:AFrep1}
	\hat {\ve{S}}_{\ve{i}} = \frac{1}{2} \ve{\hat{ f}}^{\dagger}_{\ve{i}}   \ve{\sigma}  \vec{ \hat{f}} ^{\phantom\dagger}_{\ve{i}},
\end{eqnarray}
where $\vec{\hat{f}}^{\dagger}_{\ve{i}}  \equiv  \left(\hat {f}^{\dagger}_{\ve{i},\uparrow}, \hat f^{\dagger}_{\ve{i},\downarrow} \right) $ is a two-component fermion on site $\ve{i}$ with the local constraint 
\begin{eqnarray} \label{Eq:AFrep2}
	\hat{n}_{\vec{i}} = \vec{\hat{f}}^{\dagger}_{\ve{i}}  \vec{\hat{f}}^{\phantom\dagger}_{\ve{i}} = 1,
\end{eqnarray}
and $\ve{\sigma}$ corresponds to the vector of Pauli spin-1/2 matrices. The constraint in Eq.~(\ref{Eq:AFrep2}) is implemented exactly by including a  Hubbard-$U$ term on each site and taking the limit $U \rightarrow  \infty$. In the above representation the Hamiltonian (\ref{eq:spin_peierls_hamiltonian}) that we will simulate reads,
\begin{eqnarray} \label{eq:ham_qmc}
	\hat{H}_{{\rm QMC}}   & = & \sum_{b} \left( 1 + \hat{Q}_{b} \right) \Biggl[ J_b \left( \frac{1}{4} - \frac{1}{4} \left(\hat{K}_b\right)^2 \right)  \nonumber \\ 
	&&+  \left( \sum_{\alpha, \beta} \frac{|\Gamma_{b}^{\alpha,\beta} |s^{\alpha, \beta} }{2} \left(\hat{K}_{b}^{\alpha, \beta}\right)^2 - C_b \right) \Biggr] \nonumber \\
	&&+ \sum_b \left( \frac{\hat{P}_{b}^2}{2m} + \frac{k}{2} \hat{Q}_{b}^2 \right) + \frac{U}{2}  \sum_{\ve{i}} \left(\hat{n}_{\ve{i}} - 1 \right)^2
\end{eqnarray}
where \(\hat{K}_b\) is the kinetic energy operator,
\begin{eqnarray} 
	\label{eq:gauge}
	\hat{K}_{b}^{\alpha, \beta} = \hat{S}_{\ve{i}}^{\alpha}  &+& \frac{\Gamma_{b}^{\alpha,\beta}s^{\alpha, \beta}} {|\Gamma_{b}^{\alpha,\beta} |}   \hat{S}_{\ve{j}}^{\beta} , \quad C_{b}=\sum_{\alpha, \beta }\frac{|\Gamma_{b}^{\alpha,\beta} |s^{\alpha, \beta}}{4} 
\end{eqnarray}
and \( s^{\alpha, \beta} = \pm 1 \) \cite{SatoT21}. In this framework, the operator \(\left(\hat{n}_{\ve{i}} - 1 \right)^2\) commutes with the Hamiltonian $\hat H_{\rm{QMC}}$, indicating that the $\vec{\hat{f}}$-fermion parity $ (-1)^{\hat{n}_{\ve{i}} }$ remains locally conserved. Due to this symmetry characteristic, introducing a positive Hubbard-$U$ interaction term effectively enforces a projection onto the odd parity sector, where $(-1)^{\hat{n}_{\ve{i}} }   = -1$, thus imposing the constraint in Eq. (\ref{Eq:AFrep2}). Within this parity sector, the restricted Hamiltonian is given by $ \left. \hat{H}_{\rm{QMC}} \right|_{(-1)^{\hat{n}_{\ve{i}}} = -1} = \hat{H} + C$, where $C$ represents a constant.

To proceed, we carry out a Trotter decomposition and a discrete Hubbard-Stratonovitch (HS) transformation to decouple the squared fermion bilinear interaction terms in the Hamiltonian, Eq.~(\ref{eq:ham_qmc}). Using the Gauss-Hermite quadrature, we introduce three HS fields \(l_{\ve{i},\tau}\), \(l_{b, \tau}\) and \(l_{b_{\alpha, \beta}, \tau}\) which handle the dynamics of the Hubbard-\(U\) and non-frustrating and frustrating spin-spin interactions, respectively. For the phonons, we work on the position basis \( \hat{Q}_{b} \left| \phi \right> = \phi_{b} \left| \phi \right> \). The partition function can then be written as
\begin{eqnarray} \label{eq:ZQMC}
  Z &  = & \sum_{C} \int \prod_{b,\tau} d\phi_{b,\tau} \prod_{\ve{i},\tau} \gamma(l_{\ve{i},\tau}) \prod_{b,\tau} \gamma(l_{b,\tau}) \prod_{b, \alpha, \beta,\tau} \gamma(l_{b_{\alpha, \beta},\tau})  \nonumber \\
  &&\times e^{-S_{\phi}} {\rm{Tr}_F} \Bigg[  \prod_{\tau} \prod_{b} e^{\sqrt{1+\phi_{b,\tau}} \sqrt{\Delta\tau J_b / 4} \eta(l_{b, \tau}) \hat{K}_b} \nonumber \\
  && \times \prod_{\alpha,\beta}  e^{  \sqrt{1+\phi_{b,\tau}}   \sqrt{\Delta\tau|\Gamma_{b}^{\alpha,\beta} |s^{\alpha, \beta}/2 }  \eta(l_{b_{\alpha, \beta},\tau}) \hat{K}^{\alpha, \beta}_{b}} \nonumber \\
  &&\times \prod_{\ve{i}} e^{  \sqrt{-U\Delta\tau / 2} \eta(l_{\ve{i}, \tau}) (\hat{n}_{\ve{i}} - 1)} \Bigg],
\end{eqnarray}
where \( \Delta\tau \) is the Trotter step and \( L_{\rm Trotter} = \beta / \Delta\tau \) defines the number of imaginary-time slices in the Trotter decomposition. The phonon action is given by
\begin{eqnarray}
	S_{\phi} = \Delta\tau \sum_{b,\tau} \Bigg[ \frac{m}{2} && \frac{ ( \phi_{b,\tau+1}-\phi_{b,\tau})^2}{\Delta\tau^2} \nonumber \\ 
	 + \frac{k}{2}  &&\left(\phi_{b, \tau} - \frac{C_b}{k} + \frac{J_b}{4k} \right)^2 \Bigg]. 
\end{eqnarray}
Each lattice site hosts one HS field, \( l_{i,\tau} \), and each bond hosts two HS fields, \( l_{b,\tau} \) and \( l_{b_{\alpha, \beta},\tau} \), as well as one phonon field, \( \phi_{b,\tau} \) field. 
Hence, the configuration space is \( C = \{ l_{i, \tau}; l_{b,\tau}; l_{b_{\alpha, \beta},\tau}; \phi_{b,\tau} \} \). The HS fields can take discrete values \(l = \pm 2, \pm 1\) and the phonon field is a continuous field type which satisfies periodic boundary conditions in the temporal direction \( \phi_{b,L_{\rm Trotter}+1}=\phi_{b,1} \). The pre-factors \(\gamma(l)\) and \(\eta(l)\) are given by
\begin{eqnarray}
	\gamma(\pm 1) = 1 + \sqrt{6}/3, \quad \eta(\pm 1) = \pm \sqrt{2\left( 3 - \sqrt{6} \right)}, \nonumber \\
	\gamma(\pm 2) = 1 - \sqrt{6}/3, \quad \eta(\pm 2) = \pm \sqrt{2\left( 3 + \sqrt{6} \right)}. 
\end{eqnarray}

\section{Results and Discussion}
\label{sec:results}

In the following section, we present the results of our study. First, we start by comparing our approach for the Heisenberg chain with an established method for spin-boson systems, namely SSE with wormhole updates \cite{Weber21a}. Next, we show how the negative sign problem behaves for the Kitaev-Heisenberg model with phonons. We then present results of the temperature dependence of the uniform magnetic susceptibility, both with and without phonons, for the one- and two-dimensional non-frustrated Heisenberg model as well as the frustrated Kitaev-Heisenberg model. Finally, we discuss the physics of the spin-phonon coupling in these various models.

The data was obtained using the Algorithms for Lattice Fermions (ALF)  \cite{ALF_v2} implementation of the  AFQMC.  We have adopted a single spin-flip algorithm for the phonon and HS fields. In the Heisenberg simulations the number of Trotter slices used was \(L_{\text{Trotter}} = 100\). In the Kitaev-Heisenberg simulations, we have used a range of $L_{\text{Trotter}} \in [20, 200]$ depending upon the temperature.

\subsection{Comparing with SSE and Negative Sign Problem}
\label{sub_sec:compare_with_SSE}

\begin{figure}[t]              
	\centering
  \begin{tabular}{c}
    (a)    \\ 
   	\centerline{\includegraphics[width=.4\textwidth]{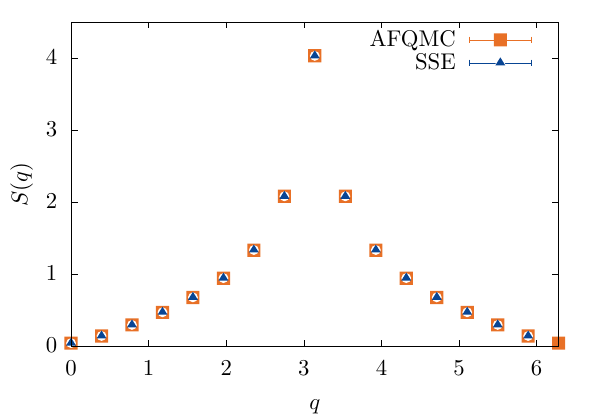}}\\
  \end{tabular}
  \begin{tabular}{c}
    (b)  \\ 
   	\centerline{\includegraphics[width=.4\textwidth]{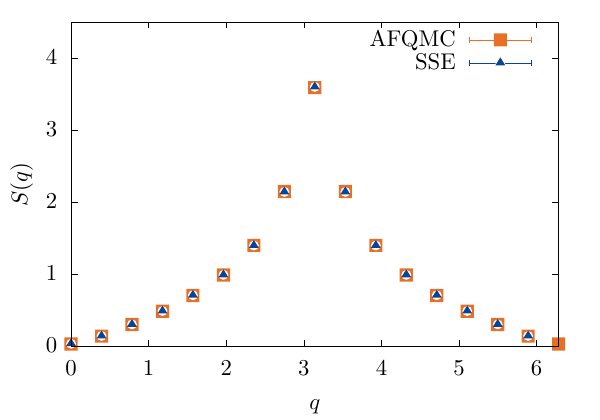}}
	\end{tabular}
	\caption{Spin structure factor as a function of lattice momentum for the one-dimensional spin-peierls Heisenberg chain with \(L = 16\) sites at temperature \(\beta J = 8\). The reference data was produced with the SSE wormhole algorithm \cite{Weber21a}. (a) shows the result for \(\lambda = 0\) and (b) for \(\lambda = 0.3\) at \(\omega_0 = 0.25\). }
	\label{fig:SSE}
\end{figure}

\begin{figure}[!t]
	\centering
  \includegraphics[width=.4\textwidth]{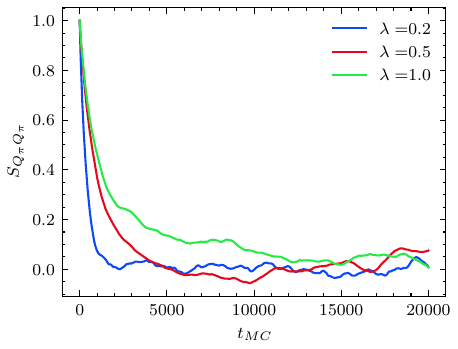}
	\caption{Autocorrelation time for the equal time correlation function of the phonon fields, at momentum \(Q = \pi\), as a function of the spin-phonon coupling constant \(\lambda\). Here, $\omega_0 = 0.5$, \(\beta = 16\) and $L = 32$ chain. 
	Here, the time unit corresponds to a single sweep in which  all the fields in the Euclidean space-time lattice  are visited once.} 
	\label{fig:auto_corr_heisenberg}                                
\end{figure}

\begin{figure}[t]
	\centering
	\includegraphics[width=0.4\textwidth]{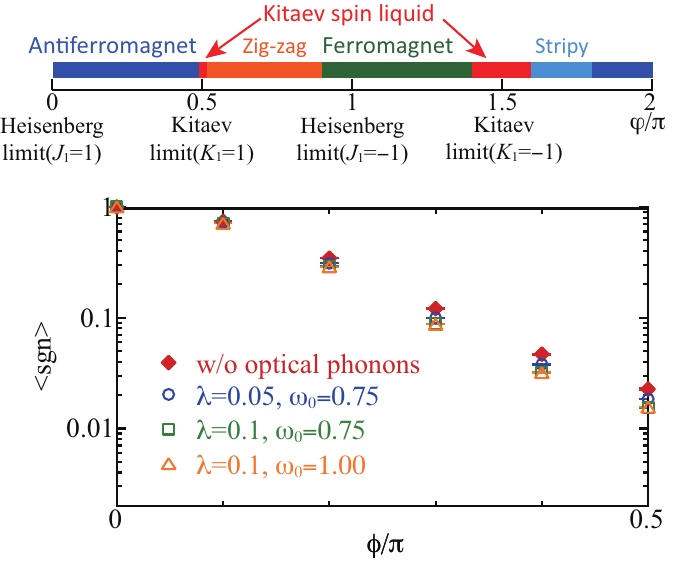}
	\caption{Proposed ground-state phase diagram  for the original Kitaev-Heisenberg model as a function of $\phi/\pi$ \cite{chaloupka_zigzag_2013}, without phonons, and the average sign of the AFQMC simulations for systems with and without phonons as a function of \(\phi/\pi\).	Here the temperature $T/A=1/1.8$ and lattice size $N=32$. }
	\label{fig:sign_kiteav_heisenberg}     
\end{figure}

\begin{figure}[!t]
	\centering
	\includegraphics[width=.4\textwidth]{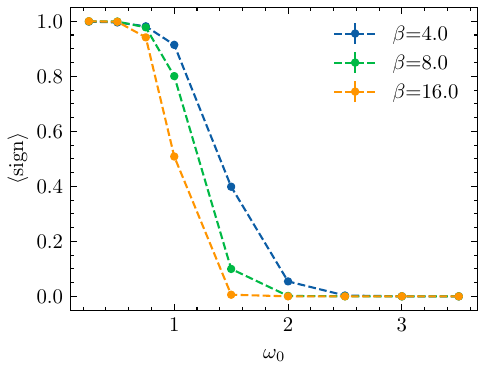}
	\caption{Average sign as a function of the phonon frequency \(\omega_0\) for different inverse temperatures and spin-phonon coupling \(\lambda = 0.2\) . Here, $L_x = L_y = 12$ square lattice.}
	\label{fig:sign_heisenberg}                                
\end{figure}

To validate our method, we compare it against an established method, the SSE wormhole algorithm \cite{Weber21a}. This is a very efficient algorithm for the simulation of spin-boson models in which the boson degrees of freedom are integrated out at the expense of a retarded spin-spin interaction. As the SSE method does not rely on a fermionic representation of the spins, the sampled configuration space is constituted by spin states which are efficiently sampled with global updates, called directed loop updates \cite{Sandvik02}. 

Figure \ref{fig:SSE} shows the momentum-resolved static spin structure factor 
\begin{eqnarray} \label{eq:static_structure_factor}
	S(\ve{q}) = \left< \vec{\hat{S}}_{\ve{q}} \cdot \vec{\hat{S}}_{- \ve{q}} \right>
\end{eqnarray}
computed with both SSE and AFQMC methods, for the Heisenberg chain with \(L = 16\) sites at inverse temperature \(\beta J = 8\) for \(\lambda = 0\) and \(\lambda = 0.3\) with \(\omega_0 = 0.25\). The results of both methods  agree well, however we cannot stress enough the fundamental differences between both approaches. For spin-phonon Hamiltonians where it is possible to find a formulation without the negative sign problem, the SSE wormhole algorithm is undoubtably a better choice to tackle the problem. 

To highlight this, in Fig. \ref{fig:auto_corr_heisenberg} we show the autocorrelation times for the equal-time phonon correlation function at \(Q = \pi\) computed with our AFQMC method. We see that the autocorrelation time either for the spin or phonon variables are of the order of \(10^3-10^4\) sweeps depending upon electron phonon coupling. This originates from the fact that 
we  sample the phonon fields  with a single spin flip algorithm.  On the other hand, it is well documented that for SSE with directed loop updates the autocorrelation times are very small \cite{Sandvik02,Weber21a}. We note that one could use global updates for the phonon and HS fields to speed up the simulation. In particular, Langevin \cite{Goetz21} or Hybrid Monte Carlo  \cite{Beyl17} updates prove efficient to propose global moves on a given subset of fields \cite{Cheng24}.

In the Kitaev-Heisenberg model, where frustration is present for any finite Kitaev coupling, Monte Carlo sampling is hindered by the notorious negative sign problem \cite{troyer_computational_2005,SatoT21}. This problem essentially leads to an exponential decrease in the efficiency of Monte Carlo sampling. Its severity is measured through the average value of the sign, defined as
\begin{eqnarray}
	\left< \text{sign} \right> = \left< \frac{\text{Re}\left(e^{-S(C)}\right)}{\left| \text{Re}\left(e^{-S(C)}\right) \right|} \right>,
\end{eqnarray}
where \(S(C)\) represents the action for a configuration \(C\), such as in Eq. \eqref{eq:ZQMC}. We utilise the fact that the partition function $Z  =  \sum_{C} e^{-S(C)} $ is real, allowing us to sample $\text{Re}\left( e^{-S(C)}\right) $. We refer the  reader to Ref.~\cite{ALF_v2} for further details. For SSE-based methods, the sign problem is particularly severe, rendering simulations quickly impractical \cite{weber_quantum_2022, hangleiter_easing_2020, alet_sign-problem-free_2016, reingruber_thermodynamics_2024}. With our method, simulations of the Kitaev-Heisenberg model are possible to temperatures up to approximately twice the magnetic scale \(T/A = 1/1.8\) \cite{SatoT21}. Figure \ref{fig:sign_kiteav_heisenberg} shows the proposed ground-state phase diagram of the Kitaev-Heisenberg model \cite{chaloupka_zigzag_2013}, without phonons, and the average sign as a function of \(\phi/\pi\). As we deviate from the Heisenberg point at \(\phi/\pi = 0\), the average sign decreases due to the increase in the Kitaev coupling strength. Remarkably, the introduction of bond phonons at low spin-phonon coupling \(\lambda < 0.2\) does not significantly worsen the negative sign problem. This allows us to study the effects of spin-phonon coupling exactly at temperatures twice as low as the magnetic energy scale, even in the Kitaev limit \(\phi/\pi = 0.5\).

The above is valid  in the limit where \(\sqrt{1 + \phi_{b, \tau}}\) remains real. Fo a fixed spin-phonon coupling, increasing the phonon frequency results in large amplitude fluctuations of the $\phi_{b,\tau}$ field. In this limit a negative sign problem will arise in the action Eq. \eqref{eq:ZQMC} since the factor \(\sqrt{1 + \phi_{b, \tau}}\) becomes imaginary. This behaviour is explicitly shown for the  Heisenberg model in Fig. \ref{fig:sign_heisenberg} where the sign problem develops as the  phonon frequency is increased. 

\subsection{Uniform magnetic susceptibility}
\label{sub_sec:uniform_spin_sus}

\begin{figure}[t]
	\centering
	\begin{tabular}{c}
    \hspace{2pt} (a) \\ 
    \hspace{-9pt} \includegraphics[width=.47\textwidth]{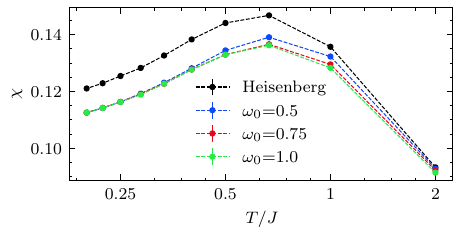}
  \end{tabular}
  \begin{tabular}{c}
    (b) \\ 
   	\centerline{\includegraphics[width=.45\textwidth]{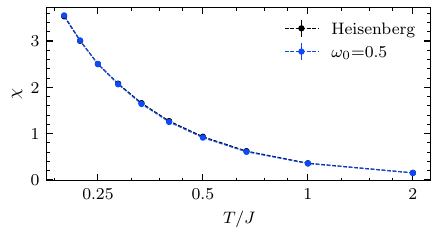}}
	 \end{tabular}
	\caption{Temperature dependence  of the uniform magnetic susceptibility for the antiferromagnetic (a) and ferromagnetic (b) Heisenberg model on a one-dimensional chain with and without phonons at different values of \(\omega_0\). Here, $\lambda = 0.2$ and $L_x = 32$. The black line corresponds to the plain Heisenberg model without phonons.}
	\label{fig:mag_suscep_1D}
\end{figure}

\begin{figure}[t]
	\centering	
  \begin{tabular}{c}
    \hspace{2pt} (a) \\ 
   	\hspace{-9pt} \includegraphics[width=.47\textwidth]{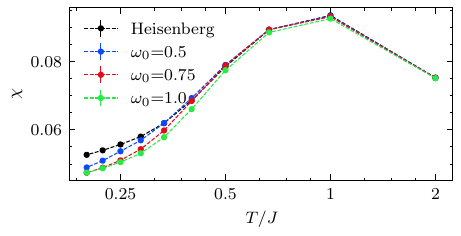}
  \end{tabular}
  \begin{tabular}{c}
    (b) \\ 
   	\centerline{\includegraphics[width=.45\textwidth]{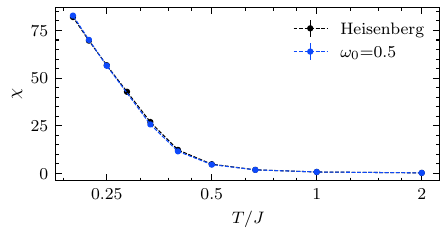}}
	 \end{tabular}
	\caption{Temperature dependence of the uniform magnetic susceptibility for the antiferromagnetic (a) and ferromagnetic (b) Heisenberg model on a square lattice with phonons at different values of \(\omega_0\). Here, $\lambda = 0.2$, $L_x = L_y = 16$ square lattice. The black line corresponds to the plain Heisenberg model without phonons. }
	\label{fig:mag_suscep_2D}
\end{figure}

\begin{figure}[t]
	\centering	
  \includegraphics[width=.475\textwidth]{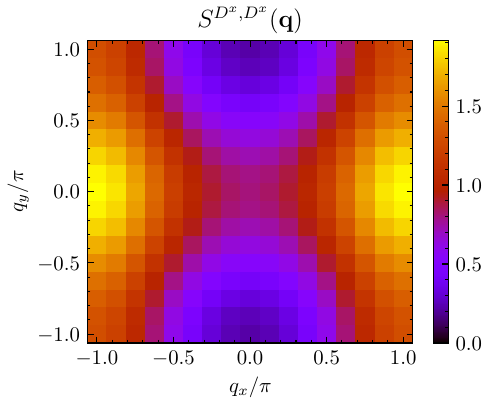}
	\caption{Momentum resolved static dimer structure factor for the antiferromagnetic Heisenberg model on a square lattice \(L_x = L_y = 16\) at inverse temperature \(\beta J = 5\) and \(\lambda = 0.2\), \(\omega_0 = 0.5\). The dimer structure factor is given by \(S^{D_x,D_x} = \left< \hat{D}^x_{\ve{q}} \hat{D}^x_{- \ve{q}} \right>\) where \(\hat{D}^x_{\ve{i}} = \ve{\hat{S}}_{\ve{i}} \cdot \ve{\hat{S}}_{\ve{i} + \ve{x}}\). }
	\label{fig:vbs_2D}
\end{figure}

\begin{figure*}[!t]
	\centering
	\includegraphics[width=0.95\textwidth]{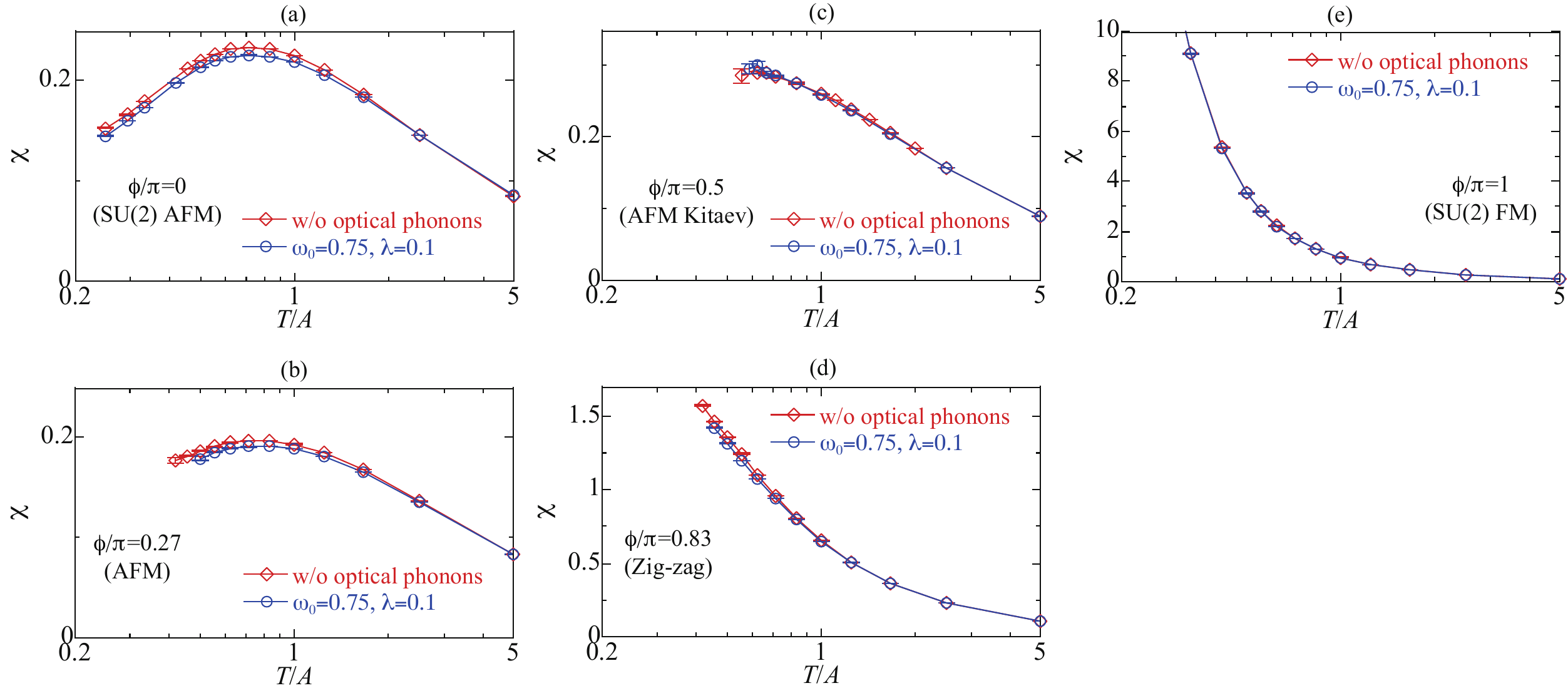}
	\caption{Temperature dependence  of the uniform magnetic susceptibility for the Kitaev-Heisenberg model with phonons at different values of $\phi/\pi$. (a) \(\phi/\pi = 0\) (\(\text{SU}(2)\) symmetric antiferromagnetic Heisenberg model), (b) \(\phi/\pi = 0.27\) (\(\text{SU}(2)\) symmetry is broken by a finite Kitaev coupling), (c) \(\phi/\pi = 0.5\) (antiferromagnetic Kitaev model), (d) \(\phi/\pi = 0.83\) (ground state has zig-zag ordering), and (e) \(\phi/\pi = 1\) (\(\text{SU}(2)\) symmetric ferromagnetic Heisenberg model). Here, $\lambda=0.1$, \(\omega_0 = 0.75\) and $N=32$ lattice.  
	}
	\label{fig:mag_suscep_kitaev}
\end{figure*}

To briefly discuss the physics in our simulations, we study the effects of spin-phonon coupling on the temperature dependence of the uniform magnetic susceptibility, 
\begin{eqnarray}
	\chi = \frac{1}{N} \int_{0}^{\beta} d\tau \left< \hat{\ve{S}}(\tau) \cdot  \hat{\ve{S}}(0) \right>,
\end{eqnarray}
where \(\hat{\ve{S}} = \sum_{\ve{i}} \hat{\ve{S}}_{\ve{i}}\) denotes the total spin operator and \(N\) signifies the number of unit cells. The uniform magnetic susceptibility provides insight into the available spin excitations as a function of temperature. For the one-dimensional antiferromagnetic chain these excitations are spinons, fractionalised particles that carry spin-\(1/2\) and no charge. For two-dimensional antiferromagnets these excitations are spin-waves, otherwise known as magnons. For the Kitaev model, the excitations are fractionalised in the form of \(\mathbb{Z}_2\) fluxes and Majorana fermions \cite{kitaev_anyons_2006}. 

In the adiabatic limit \(\omega_0 \to 0\), one can neglect the dynamics of the phonons and treat them as lattice distortions. It is then expected that antiferromagnetic systems, at zero temperature, develop a transition to a dimerised phase, known as the spin-Peierls transition \cite{giamarchi_quantum_2003}.  This transition opens a spin gap and the  magnetic susceptibility will be activated in the low temperature limit. For finite \(\omega_0\), the problem becomes severely harder to solve. Integrating out the phonons will generate a non-local and retarded interaction between the spin dimer operator \cite{Weber17,giamarchi_quantum_2003}.

We begin our discussion with the Heisenberg model in one and two dimensions, and then turn to the Kitaev-Heisenberg model on the honeycomb lattice. Note that the results presented here are for  spin-phonon couplings ranging from \(\lambda = 0.1 \) to \(\lambda = 0.2 \).

\subsubsection{Heisenberg Model in One-Dimension}

Figure \ref{fig:mag_suscep_1D} shows the magnetic susceptibility as a function of temperature for the Heisenberg chain without (black line) and with phonons at spin-phonon coupling \(\lambda = 0.2\) for different phonon frequencies. The results are presented for antiferromagnetic (a) and ferromagnetic (b) interactions. Immediately we see a noticeable difference between the two different couplings; the ferromagnetic chain does not change its behaviour in the presence of phonons, whereas the antiferromagnetic chain exhibits a significant reduction in magnetic susceptibility due to the phonon coupling.

The results for the Heisenberg chain can be readily understood within the framework of bosonisation \cite{giamarchi_quantum_2003}. In bosonisation we represent the low-energy physics of the Heisenberg chain by an effective continuum theory of bosons, also known as a Sine-Gordon model. This model consists of a free part and a umklapp term. One of the relevant parameters in this theory is the Luttinger Liquid parameter \(K\). The antiferromagnetic case corresponds to \(K = 1/2\) whilst \(K \to \infty\) to the ferromagnetic case. 

Under the aforementioned adiabatic limit, the exchange constant \(J\) can be modulated such that 
\begin{eqnarray}
	J \to J + \delta J (-1)^{i},
\end{eqnarray}
where \(i\) labels a site on the chain. Under these conditions, the relevance of this periodic perturbation can be studied using renormalisation theory. It turns out to be relevant if \(4 - 2K \geq 0\) \cite{giamarchi_quantum_2003}. In the ferromagnetic regime, this perturbation is highly irrelevant, so the phonons do not couple to the spins, as seen in Fig. \ref{fig:mag_suscep_1D} (b). By contrast, in the antiferromagnetic regime, the perturbation is relevant, implying that at zero temperature the magnetic susceptibility tends to vanish. In our simulations, as shown in Fig. \ref{fig:mag_suscep_1D} (a), we do not fully capture this behaviour because of our relatively high temperature, but we do observe the precursors of a gap opening. In particular, we expect that spectral weight is shifted from lower to higher frequencies as the temperature is lowered, thus resulting in an overall reduction of the uniform spin susceptibility. To confirm whether these statements also hold for finite phonon frequencies, we can investigate the behaviour in the anti-adiabatic regime, this is \(\omega_0 \to \infty\). In this limit, the coupling of the umklapp term in the Sine-Gordon action is renormalised as \cite{giamarchi_quantum_2003, hofmeier_spin-peierls_2024}
\begin{eqnarray}
	\frac{2J}{(2\pi)^2} \to \frac{2J}{(2\pi)^2} - \lambda.
\end{eqnarray}
If the spin-phonon coupling \(\lambda\) is ufficiently large to make the umklapp term negative, this term becomes irrelevant, and there will be a spin-Peierls transition for the antiferromagnetic coupling. For ferromagnetic couplings, this term is already irrelevant, so no spin-Peierls transition is observed. This can be observed in Fig. \ref{fig:mag_suscep_1D}. 
Our findings in both regimes thus coincide with the numerical results obtained at finite phonon frequencies and relatively high temperatures.

\subsubsection{Heisenberg Model in Two-Dimensions}
We now consider the same Heisenberg model but on a two-dimensional square lattice. Figure \ref{fig:mag_suscep_2D} presents the temperature dependence of the magnetic susceptibility for the antiferromagnetic (a) and ferromagnetic (b) Heisenberg models on the square lattice with \(\lambda = 0.2\) and various \(\omega_0\) values. We again observe that, in the ferromagnetic case, the susceptibility remains unchanged upon the inclusion of phonons, implying that the phonons do not couple to the spins in this regime.  By contrast, in the antiferromagnetic case, we still observe a reduction  of the magnetic susceptibility. 

To explain the results observed in Fig. \ref{fig:mag_suscep_2D} we use a parton mean field theory of the Heisenberg model \cite{saremi_quantum_2007}. 
Even though the excitations in this case are not explicitly fractionalised, Ref.~\cite{dalla_piazza_fractional_2015} identifies fractional quasiparticles (in the form of deconfined spinons) in the high energy spin spectrum of \(\text{Cu(DCOO)}_2 4\text{D}_{2}\text{O}\), a known realisation of the antiferromagnetic Heisenberg model on the square lattice.
The existence and nature of these quasiparticles are still under debate \cite{shao_nearly_2017, powalski_mutually_2018}. 
In Ref.~\cite{powalski_mutually_2018} the authors argue that the features of the high-energy spectrum observed experimentally are accounted for in linear spin wave theory with attractive magnon-magnon interactions, whilst in Ref.~\cite{shao_nearly_2017} they are identified as nearly deconfined spinons, a precursor to the deconfined quantum critical point of the \(J-Q\) model \cite{sandvik_evidence_2007}. 
Ultimately, it remains plausible to interpret our data using a parton mean field theory, given the temperature ranges of our simulations. We again write our spin operators as Abrikosov fermions, Eq. \eqref{Eq:AFrep1}, and use the following mean field decoupling 
\begin{eqnarray}
\left< \vec{\hat{f}}_{\ve{i}}^\dagger \vec{\hat{f}}_{\ve{j}} \right> = \chi_{\ve{i}, \ve{j}}
\end{eqnarray}
with \( \chi_{\ve{i}, \ve{j}} \in \mathbb{C} \). To restrict the Hilbert space to single occupied states and to reproduce a ferromagnetic system we add the following constraints to the Hamiltonian as Lagrange multipliers 
\begin{eqnarray}
	\hat{H}_{\lambda_1} &=& \lambda_1 \sum_{\ve{i}} \left( \vec{\hat{f}}_{\ve{i}}^\dagger \vec{\hat{f}}_{\ve{i}} - 1 \right), \\
	\hat{H}_{\lambda_2} &=& \lambda_2 \sum_{\ve{i}} \left( \hat{f}_{\ve{i}, \uparrow}^\dagger \hat{f}_{\ve{i}, \uparrow} - \hat{f}_{\ve{i}, \downarrow}^\dagger \hat{f}_{\ve{i}, \downarrow} - 1 \right),
\end{eqnarray}
respectively. The parameter \( \lambda_1 \) acts as a chemical potential and for a half-filled system it should vanish, whilst the parameter \( \lambda_2 \) controls whether the ground state has ferromagnetic order or not such that it breaks \(\text{SU}(2)\) symmetry. 

For the antiferromagnetic case \(\lambda_2 = 0\). According to Lieb's theorem \cite{lieb_flux_1994}, the configuration of the \( \chi_{\ve{i}, \ve{j}} \in \mathbb{C} \) parameters that minimises the total energy is the one with a \(\pi\) flux per plaquette. The band structure of this system has two Dirac points at \(\ve{K}^{\pm}\) \cite{Affleck88}. At low energies the Hamiltonian can be mapped to an effective Dirac Hamiltonian, obtained by linearising the dispersion relation near the Dirac points. Here, low-lying particle-hole excitations have momentum \(\ve{Q} = \ve{K}^+ - \ve{K}^- = (\pi, 0)\), meaning that adding phonons should drive the ground state into a \((\pi, 0)\) Valence Bond Solid (VBS) phase. In turn this generates a mass term in the Dirac equation thereby opening up a gap. In our simulations, Fig. \ref{fig:mag_suscep_2D} (a), we observe a  reduction of the magnetic susceptibility upon the inclusion of phonons, which evidences the development of a gap at zero temperature. Furthermore we also observe dominant  \((\pi, 0)\) VBS fluctuations developing for \(\omega_0 = 0.5\) at inverse temperature \(\beta J = 5\) [see Fig. \ref{fig:vbs_2D}].

For the ferromagnetic case, \(\lambda_2 \neq 0\), so particle-hole symmetry is broken in each spin sector and Lieb's theorem does not apply. Adopting a mean-field ansatz with real and uniform \(\chi_{\ve{i}, \ve{j}} = \chi\), \(\chi \in \mathbb{R}\), we obtain the following energy dispersion 
\begin{eqnarray}
	\epsilon_{\ve{q},\sigma} = \left[ - J \chi \left( \cos(q_x) + \cos(q_y) \right) + \lambda_1  \right]\mathbb{I} + \sigma \lambda_2.
\end{eqnarray}
As a result, the energy spectrum is spin polarised. 
As the polarisation grows,  the phase  space  available for  phonon scattering decreases.  In particular, in the the  fully  polarised phase,  the lower band (\( \epsilon_{\ve{q}, \downarrow} \)) is completely filled whilst the upper band (\( \epsilon_{\ve{q}, \uparrow} \)) remains empty such that coupling to the phonon modes is completely suppressed. This is consistent with Fig. \ref{fig:mag_suscep_2D} (b).

\subsubsection{Kitaev-Heisenberg model}

We now turn to the results and discussion for the frustrated Kitaev-Heisenberg model on the honeycomb lattice. In Fig. \ref{fig:mag_suscep_kitaev} we present the temperature dependence of the magnetic susceptibility for different values of \(\phi/\pi\).  Unlike the previous section,  here we  use a smaller  value of the electron-phonon coupling \(\lambda = 0.1\).  We again adopt an Abrikosov representation of the spins, Eq. \eqref{Eq:AFrep1}.  Using a  Fierz identity for \(\text{SU}(2)\) and some algebraic manipulation, we arrive at 
\begin{eqnarray} \label{eq:kitaev_high_abrikosov}
	\hat{H} = - \left( \frac{K}{4} + \frac{J}{4} \right) \sum_{\left< \ve{i}, \ve{j} \right>_\alpha} \left( \hat{D}_{\ve{i}, \ve{j}}^\dagger \hat{D}_{\ve{i}, \ve{j}} + \hat{D}_{\ve{i}, \ve{j}} \hat{D}_{\ve{i}, \ve{j}}^\dagger \right) \nonumber \\
	- \frac{K}{4}  \sum_{\left< \ve{i}, \ve{j} \right>_\alpha} \left( {\left(\hat{D}_{\ve{i}, \ve{j}}^\alpha\right)}^\dagger \hat{D}_{\ve{i}, \ve{j}}^\alpha + \hat{D}_{\ve{i},\ve{j}}^\alpha {\left(\hat{D}_{\ve{i}, \ve{j}}^\alpha\right)}^\dagger \right),
\end{eqnarray}
with 
\begin{eqnarray}
	\hat{D}_{\ve{i}, \ve{j}} =  \sum_s \hat{f}_{\ve{i}, s}^\dagger \hat{f}_{\ve{j}, s}, \quad \hat{D}_{\ve{i}, \ve{j}}^\alpha =  \sum_{s, s^\prime}  \hat{f}_{\ve{i}, s}^\dagger \sigma^{\alpha}_{s, s^\prime} \hat{f}_{\ve{j}, s^\prime} \nonumber
\end{eqnarray}
serving as bond operators. This is an exact rewriting of the Kitaev-Heisenberg model  provided that the 
constraint $\sum_{s} \hat{f}_{\ve{i}, s}^\dagger \hat{f}^{}_{\ve{i}, s} = 1$  is  imposed.  The operator \(\hat{D}_{\ve{i},\ve{j}}\) accounts for the spin-independent hopping between sites \(\ve{i}\) and \(\ve{j}\), so it preserves \(\text{SU}(2)\) symmetry. On the other hand, \(\hat{D}_{i, j}^\alpha\) accounts for a spin-dependent hopping for a bond oriented in direction \(\alpha\), allowing only spins aligned with \(\alpha\) to hop. Our parametrisation reads:   $K = A \sin(\phi)$ and $J = A \cos(\phi)$, where $A$ is the magnetic energy scale. 

We begin our analysis with the \(\text{SU}(2)\) symmetric antiferromagnetic (ferromagnetic) models, \(\phi/\pi = 0\ (1)\). As observed in the Heisenberg model on the chain and on the square lattice, we find similar overall behaviour: the ferromagnetic system does not couple to the phonons and the antiferromagnetic one does. The explanation for this behaviour is quite similar to the explanation for the Heisenberg model on the square lattice. In the  parton construction and for the antiferromagnetic case,  the low lying particle-hole excitations have momentum 
\(\ve{Q} = \ve{K}^+ - \ve{K}^-\) that allow for excitations  between the Dirac cones.  At finite 
electron-phonon coupling we  expect an instability to a  Kekul\'e phase \cite{Lang13}  that  will open up a gap in the  spin sector.   The data of Fig. \ref{fig:mag_suscep_kitaev} (a) shows a  maximum at $T/A \simeq 0.7$, which corresponds to a characteristic scale at which short-ranged spin-spin correlations  emerge.  In terms of the partons this corresponds to a coherence scale. We understand the slight reduction of the spin susceptibility as a precursor to gap opening in which spectral  weight is transfered to higher frequencies. 

At the Kitaev point \(\phi/\pi = 0.5\) the model is exactly solvable without phonons \cite{kitaev_anyons_2006}. For temperatures below the flux gap, the low-energy physics of the Kitaev model is dictated by the Dirac equation. Again the low-lying particle-hole excitations have momentum \(\ve{Q} = \ve{K}^+ - \ve{K}^-\) and the addition of phonons should drive the system to a Kekul\'e phase. In this phase, the bonds in the honeycomb lattice are modulated in a way that tunnelling is allowed between the two Dirac points. This generates a mass term in the Dirac equation and at low temperature one expects the Kitaev model to develop a spin-Peierls phase. Unfortunately, these temperature ranges are inaccessible with our AFQMC approach due to the negative sign problem \cite{SatoT21,SatoT23a}. For temperatures above the flux gap, the fluxes are completely disordered and obtaining an analytical solution becomes more difficult, since one has to average over all flux configurations. To understand the behaviour of the Kitaev model at high temperatures (\(K \approx T\)) we again use the Abrikosov representation of  Eq. \eqref{Eq:AFrep1}.  In this formulation, frustration arises  via the 
\(\hat{D}_{\ve{i}, \ve{j}}^\alpha\) terms that prohibit coherent hopping of the partons: hopping is tied to the spin orientation. As apparent from Fig.~\ref{fig:mag_suscep_kitaev} (c)  the coherence temperature is reduced in comparison to the pure Heisenberg case \cite{nasu_thermal_2015}. 
This  observation is confirmed  by the magnetropic susceptibility  of Ref.~\cite{SatoT23a}, which indicates that, at our lowest accessible temperature, the system behaves like a renormalised local moment. The data of Fig.~\ref{fig:mag_suscep_kitaev} (c)  shows that  the inclusion of phonons, leaves the magnetic susceptibility  next to invariant, a fact that we interpret in terms of locality. 

Moving away   from the  Kitaev point towards the antiferromagnetic (\(\phi/\pi = 0.27\)) and zig-zag (\(\phi/\pi = 0.83\)) phases [see Fig.~\ref{fig:mag_suscep_kitaev} (b), (d)], the  suppression of the Kitaev term relative to the  Heisenberg term restores coherence.  Consequently, coupling to the phonon degrees of freedom  leads to a  slight reduction of  the uniform magnetic susceptibility.

\section{Outlook and Conclusion}
\label{sec:conclusion}

In this work we developed a new approach to simulate generic spin-Peierls Hamiltonian using the finite temperature AFQMC method. The Hamiltonian considers interacting spins on a lattice coupled to bond Einstein phonons. To formulate our AFQMC algorithm we express the spins in terms of Abrikosov fermions and impose the Hilbert space constraint via a Hubbard-\(U\) interaction on each site. The sampled configuration space consists of the HS fields, used to decouple the spin-spin and Hubbard-\(U\) interactions, and the phonon fields. For non-frustrating spin interactions our method correctly reproduces results from established approaches, such as the SSE wormhole algorithm. Due to algorithmic differences, the SSE wormhole algorithm is a superior method to study non-frustrating spin-Peierls Hamiltonians. For frustrated systems, such as the Kitaev-Heisenberg model, SSE based methods suffer from a severe negative sign problem that renders simulations unfeasible. By contrast, with our method we are able to simulate the Kitaev-Heisenberg for temperatures up to twice the magnetic scale. Furthermore, we showed that upon the inclusion of phonons, in the adiabatic regime at fixed electron-phonon coupling, the sign problem does not get significantly worse, meaning that the same temperature regime is accessible. 

To demonstrate the potential of our fermionic based AFQMC method, we applied it to the non-frustrated Heisenberg model in one and two dimensions, as well as to the frustrated Kitaev-Heisenberg model on the honeycomb lattice. We calculated the temperature dependence of the uniform magnetic susceptibility, which is used as a measure of the spin-Peierls transition. As expected, we find traces of a zero-temperature spin-Peierls transition in the one- and two-dimensional Heisenberg models. In one dimension the transition can be explained within the bosonisation framework in which a spin-Peierls transition occurs in both adiabatic and anti-adiabatic limits. In two dimensions the transition can be explained through low-lying particle-hole excitations. On the square lattice, for an antiferromagnetic coupling, these excitations have wavevector \(\ve{Q} = (\pi, 0)\) which drives the system to a \((\pi, 0)\) VBS phase at zero temperature. For a ferromagnetic system no transition is found. For the Kitaev-Heisenberg model, within the accessible temperature ranges, our simulations show that there is no reduction of the magnetic susceptibility when including phonons at the Kitaev point. By performing an exact rewriting of the Kitaev Hamiltonian we argue that due to a spin-dependent hopping the coherence temperature of the spinons is lowered and the phonons do not couple to the spins due to thermal fluctuations. For the Kitaev model this temperature is approximately \(T \approx 0.511\) \cite{nasu_thermal_2015} and lower temperatures are inaccessible with our method due to the negative sign problem.

We believe that this method can pave the way to study signatures of fractionalisation in the phonon spectral function and apply it to the model for the Kitaev quantum spin liquid candidate \(\alpha\text{-RuCl}_3\).
Clearly, the bottle-neck is still the limitation to relatively high temperatures in the presence of strong frustration.  However, we foresee that one can further optimise the negative sign problem in the  realm of the AFQMC simulations of frustrated spin systems by further optimising the action. Possible strategies include optimal manifolds associated with complexification of the action \cite{Alexandru22}, or including more degrees of freedom in the action [see. Eq.~\ref{eq:gauge}] that can be used to optimise the negative sign.  As shown here we do not expect the negative sign problem to become more severe when coupling to phonon degrees of freedom.   

\section*{Acknowledgments}

We thank M. Weber for providing the benchmark data of Fig.~\ref{fig:SSE}.  We would like to thank W. Brenig, J. Knolle,  R. Valenti and J. Willsher for  useful discussions. We equally thank E. Huffman and J. Hofmann for discussion on how to  generalise the ALF-code to include the spin-Peierls coupling. 
We  gratefully acknowledge the Gauss Centre for Supercomputing e.V. for funding this project by providing computing time on the GCS Supercomputer SUPERMUC-NG at Leibniz Supercomputing,   (project number pn73xu) as  well  as  the scientific support and HPC resources provided by the Erlangen National High Performance Computing Center (NHR@FAU) of the Friedrich-Alexander-Universit\"at Erlangen-N\"urnberg (FAU) under the NHR project b133ae. NHR funding is provided by federal and Bavarian state authorities. NHR@FAU hardware is partially funded by the German Research Foundation (DFG) -- 440719683. 
J.I thanks the DFG for financial support under the AS 120/19-1 grant (Project number, 530989922).
T.S thanks  the W\"urzburg-Dresden Cluster of Excellence on Complexity and Topology in Quantum Matter ct.qmat (EXC 2147, project-id 390858490).   F.F.A.\ acknowledge financial support from the DFG under the grant AS 120/16-1 (Project number 493886309) that is part of the collaborative research project SFB Q-M\&S funded by the Austrian Science Fund (FWF) F 86.

\bibliographystyle{apsrev4-1}
\bibliography{./fassaad.bib,./references.bib}

\begin{thebibliography}{40}%
\makeatletter
\providecommand \@ifxundefined [1]{%
 \@ifx{#1\undefined}
}%
\providecommand \@ifnum [1]{%
 \ifnum #1\expandafter \@firstoftwo
 \else \expandafter \@secondoftwo
 \fi
}%
\providecommand \@ifx [1]{%
 \ifx #1\expandafter \@firstoftwo
 \else \expandafter \@secondoftwo
 \fi
}%
\providecommand \natexlab [1]{#1}%
\providecommand \enquote  [1]{``#1''}%
\providecommand \bibnamefont  [1]{#1}%
\providecommand \bibfnamefont [1]{#1}%
\providecommand \citenamefont [1]{#1}%
\providecommand \href@noop [0]{\@secondoftwo}%
\providecommand \href [0]{\begingroup \@sanitize@url \@href}%
\providecommand \@href[1]{\@@startlink{#1}\@@href}%
\providecommand \@@href[1]{\endgroup#1\@@endlink}%
\providecommand \@sanitize@url [0]{\catcode `\\12\catcode `\$12\catcode
  `\&12\catcode `\#12\catcode `\^12\catcode `\_12\catcode `\%12\relax}%
\providecommand \@@startlink[1]{}%
\providecommand \@@endlink[0]{}%
\providecommand \url  [0]{\begingroup\@sanitize@url \@url }%
\providecommand \@url [1]{\endgroup\@href {#1}{\urlprefix }}%
\providecommand \urlprefix  [0]{URL }%
\providecommand \Eprint [0]{\href }%
\providecommand \doibase [0]{http://dx.doi.org/}%
\providecommand \selectlanguage [0]{\@gobble}%
\providecommand \bibinfo  [0]{\@secondoftwo}%
\providecommand \bibfield  [0]{\@secondoftwo}%
\providecommand \translation [1]{[#1]}%
\providecommand \BibitemOpen [0]{}%
\providecommand \bibitemStop [0]{}%
\providecommand \bibitemNoStop [0]{.\EOS\space}%
\providecommand \EOS [0]{\spacefactor3000\relax}%
\providecommand \BibitemShut  [1]{\csname bibitem#1\endcsname}%
\let\auto@bib@innerbib\@empty
\bibitem [{\citenamefont {Balents}(2010)}]{Balents10}%
  \BibitemOpen
  \bibfield  {author} {\bibinfo {author} {\bibfnamefont {L.}~\bibnamefont
  {Balents}},\ }\href {\doibase http://dx.doi.org/10.1038/nature08917}
  {\bibfield  {journal} {\bibinfo  {journal} {Nature}\ }\textbf {\bibinfo
  {volume} {464}},\ \bibinfo {pages} {199} (\bibinfo {year}
  {2010})}\BibitemShut {NoStop}%
\bibitem [{\citenamefont {Mahan}(1990)}]{Mahan90}%
  \BibitemOpen
  \bibfield  {author} {\bibinfo {author} {\bibfnamefont {G.~D.}\ \bibnamefont
  {Mahan}},\ }\href@noop {} {\emph {\bibinfo {title} {Many-Particle
  Physics}}},\ \bibinfo {edition} {2nd}\ ed.\ (\bibinfo  {publisher} {Plenum
  Press},\ \bibinfo {address} {New York},\ \bibinfo {year} {1990})\BibitemShut
  {NoStop}%
\bibitem [{\citenamefont {Sandilands}\ \emph {et~al.}(2015)\citenamefont
  {Sandilands}, \citenamefont {Tian}, \citenamefont {Plumb}, \citenamefont
  {Kim},\ and\ \citenamefont {Burch}}]{Sandilands15a}%
  \BibitemOpen
  \bibfield  {author} {\bibinfo {author} {\bibfnamefont {L.~J.}\ \bibnamefont
  {Sandilands}}, \bibinfo {author} {\bibfnamefont {Y.}~\bibnamefont {Tian}},
  \bibinfo {author} {\bibfnamefont {K.~W.}\ \bibnamefont {Plumb}}, \bibinfo
  {author} {\bibfnamefont {Y.-J.}\ \bibnamefont {Kim}}, \ and\ \bibinfo
  {author} {\bibfnamefont {K.~S.}\ \bibnamefont {Burch}},\ }\href {\doibase
  10.1103/PhysRevLett.114.147201} {\bibfield  {journal} {\bibinfo  {journal}
  {Phys. Rev. Lett.}\ }\textbf {\bibinfo {volume} {114}},\ \bibinfo {pages}
  {147201} (\bibinfo {year} {2015})}\BibitemShut {NoStop}%
\bibitem [{\citenamefont {Blankenbecler}\ \emph {et~al.}(1981)\citenamefont
  {Blankenbecler}, \citenamefont {Scalapino},\ and\ \citenamefont
  {Sugar}}]{Blankenbecler81}%
  \BibitemOpen
  \bibfield  {author} {\bibinfo {author} {\bibfnamefont {R.}~\bibnamefont
  {Blankenbecler}}, \bibinfo {author} {\bibfnamefont {D.~J.}\ \bibnamefont
  {Scalapino}}, \ and\ \bibinfo {author} {\bibfnamefont {R.~L.}\ \bibnamefont
  {Sugar}},\ }\href {\doibase 10.1103/PhysRevD.24.2278} {\bibfield  {journal}
  {\bibinfo  {journal} {Phys. Rev. D}\ }\textbf {\bibinfo {volume} {24}},\
  \bibinfo {pages} {2278} (\bibinfo {year} {1981})}\BibitemShut {NoStop}%
\bibitem [{\citenamefont {White}\ \emph {et~al.}(1989)\citenamefont {White},
  \citenamefont {Scalapino}, \citenamefont {Sugar}, \citenamefont {Loh},
  \citenamefont {Gubernatis},\ and\ \citenamefont {Scalettar}}]{White89}%
  \BibitemOpen
  \bibfield  {author} {\bibinfo {author} {\bibfnamefont {S.}~\bibnamefont
  {White}}, \bibinfo {author} {\bibfnamefont {D.}~\bibnamefont {Scalapino}},
  \bibinfo {author} {\bibfnamefont {R.}~\bibnamefont {Sugar}}, \bibinfo
  {author} {\bibfnamefont {E.}~\bibnamefont {Loh}}, \bibinfo {author}
  {\bibfnamefont {J.}~\bibnamefont {Gubernatis}}, \ and\ \bibinfo {author}
  {\bibfnamefont {R.}~\bibnamefont {Scalettar}},\ }\href {\doibase
  10.1103/PhysRevB.40.506} {\bibfield  {journal} {\bibinfo  {journal} {Phys.
  Rev. B}\ }\textbf {\bibinfo {volume} {40}},\ \bibinfo {pages} {506} (\bibinfo
  {year} {1989})}\BibitemShut {NoStop}%
\bibitem [{\citenamefont {Assaad}\ and\ \citenamefont
  {Evertz}(2008)}]{Assaad08_rev}%
  \BibitemOpen
  \bibfield  {author} {\bibinfo {author} {\bibfnamefont {F.}~\bibnamefont
  {Assaad}}\ and\ \bibinfo {author} {\bibfnamefont {H.}~\bibnamefont
  {Evertz}},\ }in\ \href {\doibase 10.1007/978-3-540-74686-7_10} {\emph
  {\bibinfo {booktitle} {Computational Many-Particle Physics}}},\ \bibinfo
  {series} {Lecture Notes in Physics}, Vol.\ \bibinfo {volume} {739},\ \bibinfo
  {editor} {edited by\ \bibinfo {editor} {\bibfnamefont {H.}~\bibnamefont
  {Fehske}}, \bibinfo {editor} {\bibfnamefont {R.}~\bibnamefont {Schneider}}, \
  and\ \bibinfo {editor} {\bibfnamefont {A.}~\bibnamefont {Wei{\ss}e}}}\
  (\bibinfo  {publisher} {Springer},\ \bibinfo {address} {Berlin Heidelberg},\
  \bibinfo {year} {2008})\ pp.\ \bibinfo {pages} {277--356}\BibitemShut
  {NoStop}%
\bibitem [{\citenamefont {Assaad}\ \emph {et~al.}(2022)\citenamefont {Assaad},
  \citenamefont {Bercx}, \citenamefont {Goth}, \citenamefont {G{\"o}tz},
  \citenamefont {Hofmann}, \citenamefont {Huffman}, \citenamefont {Liu},
  \citenamefont {Toldin}, \citenamefont {Portela},\ and\ \citenamefont
  {Schwab}}]{ALF_v2}%
  \BibitemOpen
  \bibfield  {author} {\bibinfo {author} {\bibfnamefont {F.~F.}\ \bibnamefont
  {Assaad}}, \bibinfo {author} {\bibfnamefont {M.}~\bibnamefont {Bercx}},
  \bibinfo {author} {\bibfnamefont {F.}~\bibnamefont {Goth}}, \bibinfo {author}
  {\bibfnamefont {A.}~\bibnamefont {G{\"o}tz}}, \bibinfo {author}
  {\bibfnamefont {J.~S.}\ \bibnamefont {Hofmann}}, \bibinfo {author}
  {\bibfnamefont {E.}~\bibnamefont {Huffman}}, \bibinfo {author} {\bibfnamefont
  {Z.}~\bibnamefont {Liu}}, \bibinfo {author} {\bibfnamefont {F.~P.}\
  \bibnamefont {Toldin}}, \bibinfo {author} {\bibfnamefont {J.~S.~E.}\
  \bibnamefont {Portela}}, \ and\ \bibinfo {author} {\bibfnamefont
  {J.}~\bibnamefont {Schwab}},\ }\href {\doibase 10.21468/SciPostPhysCodeb.1}
  {\bibfield  {journal} {\bibinfo  {journal} {SciPost Phys. Codebases}\ ,\
  \bibinfo {pages} {1}} (\bibinfo {year} {2022})}\BibitemShut {NoStop}%
\bibitem [{\citenamefont {Sylju\aa{}sen}\ and\ \citenamefont
  {Sandvik}(2002)}]{Sandvik02}%
  \BibitemOpen
  \bibfield  {author} {\bibinfo {author} {\bibfnamefont {O.}~\bibnamefont
  {Sylju\aa{}sen}}\ and\ \bibinfo {author} {\bibfnamefont {A.}~\bibnamefont
  {Sandvik}},\ }\href {\doibase 10.1103/PhysRevE.66.046701} {\bibfield
  {journal} {\bibinfo  {journal} {Phys. Rev. E}\ }\textbf {\bibinfo {volume}
  {66}},\ \bibinfo {pages} {046701} (\bibinfo {year} {2002})}\BibitemShut
  {NoStop}%
\bibitem [{\citenamefont {Weber}\ \emph {et~al.}(2017)\citenamefont {Weber},
  \citenamefont {Assaad},\ and\ \citenamefont {Hohenadler}}]{Weber17}%
  \BibitemOpen
  \bibfield  {author} {\bibinfo {author} {\bibfnamefont {M.}~\bibnamefont
  {Weber}}, \bibinfo {author} {\bibfnamefont {F.~F.}\ \bibnamefont {Assaad}}, \
  and\ \bibinfo {author} {\bibfnamefont {M.}~\bibnamefont {Hohenadler}},\
  }\href {\doibase 10.1103/PhysRevLett.119.097401} {\bibfield  {journal}
  {\bibinfo  {journal} {Phys. Rev. Lett.}\ }\textbf {\bibinfo {volume} {119}},\
  \bibinfo {pages} {097401} (\bibinfo {year} {2017})}\BibitemShut {NoStop}%
\bibitem [{\citenamefont {Weber}(2022)}]{Weber21a}%
  \BibitemOpen
  \bibfield  {author} {\bibinfo {author} {\bibfnamefont {M.}~\bibnamefont
  {Weber}},\ }\href {\doibase 10.1103/PhysRevB.105.165129} {\bibfield
  {journal} {\bibinfo  {journal} {Phys. Rev. B}\ }\textbf {\bibinfo {volume}
  {105}},\ \bibinfo {pages} {165129} (\bibinfo {year} {2022})}\BibitemShut
  {NoStop}%
\bibitem [{\citenamefont {Sato}\ and\ \citenamefont {Assaad}(2021)}]{SatoT21}%
  \BibitemOpen
  \bibfield  {author} {\bibinfo {author} {\bibfnamefont {T.}~\bibnamefont
  {Sato}}\ and\ \bibinfo {author} {\bibfnamefont {F.~F.}\ \bibnamefont
  {Assaad}},\ }\href {\doibase 10.1103/PhysRevB.104.L081106} {\bibfield
  {journal} {\bibinfo  {journal} {Phys. Rev. B}\ }\textbf {\bibinfo {volume}
  {104}},\ \bibinfo {pages} {L081106} (\bibinfo {year} {2021})}\BibitemShut
  {NoStop}%
\bibitem [{\citenamefont {Sato}\ \emph {et~al.}(2024)\citenamefont {Sato},
  \citenamefont {Ramshaw}, \citenamefont {Modic},\ and\ \citenamefont
  {Assaad}}]{SatoT23a}%
  \BibitemOpen
  \bibfield  {author} {\bibinfo {author} {\bibfnamefont {T.}~\bibnamefont
  {Sato}}, \bibinfo {author} {\bibfnamefont {B.~J.}\ \bibnamefont {Ramshaw}},
  \bibinfo {author} {\bibfnamefont {K.~A.}\ \bibnamefont {Modic}}, \ and\
  \bibinfo {author} {\bibfnamefont {F.~F.}\ \bibnamefont {Assaad}},\ }\href
  {\doibase 10.1103/PhysRevB.110.L201114} {\bibfield  {journal} {\bibinfo
  {journal} {Phys. Rev. B}\ }\textbf {\bibinfo {volume} {110}},\ \bibinfo
  {pages} {L201114} (\bibinfo {year} {2024})}\BibitemShut {NoStop}%
\bibitem [{\citenamefont {Stolpp}\ \emph {et~al.}(2021)\citenamefont {Stolpp},
  \citenamefont {K{\"o}hler}, \citenamefont {Manmana}, \citenamefont
  {Jeckelmann}, \citenamefont {Heidrich-Meisner},\ and\ \citenamefont
  {Paeckel}}]{Stolpp21}%
  \BibitemOpen
  \bibfield  {author} {\bibinfo {author} {\bibfnamefont {J.}~\bibnamefont
  {Stolpp}}, \bibinfo {author} {\bibfnamefont {T.}~\bibnamefont {K{\"o}hler}},
  \bibinfo {author} {\bibfnamefont {S.~R.}\ \bibnamefont {Manmana}}, \bibinfo
  {author} {\bibfnamefont {E.}~\bibnamefont {Jeckelmann}}, \bibinfo {author}
  {\bibfnamefont {F.}~\bibnamefont {Heidrich-Meisner}}, \ and\ \bibinfo
  {author} {\bibfnamefont {S.}~\bibnamefont {Paeckel}},\ }\href {\doibase
  https://doi.org/10.1016/j.cpc.2021.108106} {\bibfield  {journal} {\bibinfo
  {journal} {Computer Physics Communications}\ }\textbf {\bibinfo {volume}
  {269}},\ \bibinfo {pages} {108106} (\bibinfo {year} {2021})}\BibitemShut
  {NoStop}%
\bibitem [{\citenamefont {Shi}\ and\ \citenamefont {Zhang}(2016)}]{Hao16}%
  \BibitemOpen
  \bibfield  {author} {\bibinfo {author} {\bibfnamefont {H.}~\bibnamefont
  {Shi}}\ and\ \bibinfo {author} {\bibfnamefont {S.}~\bibnamefont {Zhang}},\
  }\href {\doibase 10.1103/PhysRevE.93.033303} {\bibfield  {journal} {\bibinfo
  {journal} {Phys. Rev. E}\ }\textbf {\bibinfo {volume} {93}},\ \bibinfo
  {pages} {033303} (\bibinfo {year} {2016})}\BibitemShut {NoStop}%
\bibitem [{\citenamefont {Ulybyshev}\ and\ \citenamefont
  {Assaad}(2022)}]{Ulybyshev21_1}%
  \BibitemOpen
  \bibfield  {author} {\bibinfo {author} {\bibfnamefont {M.}~\bibnamefont
  {Ulybyshev}}\ and\ \bibinfo {author} {\bibfnamefont {F.}~\bibnamefont
  {Assaad}},\ }\href {\doibase 10.1103/PhysRevE.106.025318} {\bibfield
  {journal} {\bibinfo  {journal} {Phys. Rev. E}\ }\textbf {\bibinfo {volume}
  {106}},\ \bibinfo {pages} {025318} (\bibinfo {year} {2022})}\BibitemShut
  {NoStop}%
\bibitem [{\citenamefont {G\"otz}\ \emph {et~al.}(2022)\citenamefont {G\"otz},
  \citenamefont {Beyl}, \citenamefont {Hohenadler},\ and\ \citenamefont
  {Assaad}}]{Goetz21}%
  \BibitemOpen
  \bibfield  {author} {\bibinfo {author} {\bibfnamefont {A.}~\bibnamefont
  {G\"otz}}, \bibinfo {author} {\bibfnamefont {S.}~\bibnamefont {Beyl}},
  \bibinfo {author} {\bibfnamefont {M.}~\bibnamefont {Hohenadler}}, \ and\
  \bibinfo {author} {\bibfnamefont {F.~F.}\ \bibnamefont {Assaad}},\ }\href
  {\doibase 10.1103/PhysRevB.105.085151} {\bibfield  {journal} {\bibinfo
  {journal} {Phys. Rev. B}\ }\textbf {\bibinfo {volume} {105}},\ \bibinfo
  {pages} {085151} (\bibinfo {year} {2022})}\BibitemShut {NoStop}%
\bibitem [{\citenamefont {Assaad}\ and\ \citenamefont {Lang}(2007)}]{Assaad07}%
  \BibitemOpen
  \bibfield  {author} {\bibinfo {author} {\bibfnamefont {F.~F.}\ \bibnamefont
  {Assaad}}\ and\ \bibinfo {author} {\bibfnamefont {T.~C.}\ \bibnamefont
  {Lang}},\ }\href {\doibase 10.1103/PhysRevB.76.035116} {\bibfield  {journal}
  {\bibinfo  {journal} {Phys. Rev. B}\ }\textbf {\bibinfo {volume} {76}},\
  \bibinfo {pages} {035116} (\bibinfo {year} {2007})}\BibitemShut {NoStop}%
\bibitem [{\citenamefont {Karakuzu}\ \emph {et~al.}(2018)\citenamefont
  {Karakuzu}, \citenamefont {Seki},\ and\ \citenamefont
  {Sorella}}]{Karakuzu18}%
  \BibitemOpen
  \bibfield  {author} {\bibinfo {author} {\bibfnamefont {S.}~\bibnamefont
  {Karakuzu}}, \bibinfo {author} {\bibfnamefont {K.}~\bibnamefont {Seki}}, \
  and\ \bibinfo {author} {\bibfnamefont {S.}~\bibnamefont {Sorella}},\ }\href
  {\doibase 10.1103/PhysRevB.98.201108} {\bibfield  {journal} {\bibinfo
  {journal} {Phys. Rev. B}\ }\textbf {\bibinfo {volume} {98}},\ \bibinfo
  {pages} {201108} (\bibinfo {year} {2018})}\BibitemShut {NoStop}%
\bibitem [{\citenamefont {G\"otz}\ \emph {et~al.}(2024)\citenamefont {G\"otz},
  \citenamefont {Hohenadler},\ and\ \citenamefont {Assaad}}]{Goetz23}%
  \BibitemOpen
  \bibfield  {author} {\bibinfo {author} {\bibfnamefont {A.}~\bibnamefont
  {G\"otz}}, \bibinfo {author} {\bibfnamefont {M.}~\bibnamefont {Hohenadler}},
  \ and\ \bibinfo {author} {\bibfnamefont {F.~F.}\ \bibnamefont {Assaad}},\
  }\href {\doibase 10.1103/PhysRevB.109.195154} {\bibfield  {journal} {\bibinfo
   {journal} {Phys. Rev. B}\ }\textbf {\bibinfo {volume} {109}},\ \bibinfo
  {pages} {195154} (\bibinfo {year} {2024})}\BibitemShut {NoStop}%
\bibitem [{\citenamefont {Chaloupka}\ \emph {et~al.}(2013)\citenamefont
  {Chaloupka}, \citenamefont {Jackeli},\ and\ \citenamefont
  {Khaliullin}}]{chaloupka_zigzag_2013}%
  \BibitemOpen
  \bibfield  {author} {\bibinfo {author} {\bibfnamefont {J.}~\bibnamefont
  {Chaloupka}}, \bibinfo {author} {\bibfnamefont {G.}~\bibnamefont {Jackeli}},
  \ and\ \bibinfo {author} {\bibfnamefont {G.}~\bibnamefont {Khaliullin}},\
  }\href {\doibase 10.1103/PhysRevLett.110.097204} {\bibfield  {journal}
  {\bibinfo  {journal} {Physical Review Letters}\ }\textbf {\bibinfo {volume}
  {110}},\ \bibinfo {pages} {097204} (\bibinfo {year} {2013})},\ \bibinfo
  {note} {publisher: American Physical Society}\BibitemShut {NoStop}%
\bibitem [{\citenamefont {Beyl}\ \emph {et~al.}(2018)\citenamefont {Beyl},
  \citenamefont {Goth},\ and\ \citenamefont {Assaad}}]{Beyl17}%
  \BibitemOpen
  \bibfield  {author} {\bibinfo {author} {\bibfnamefont {S.}~\bibnamefont
  {Beyl}}, \bibinfo {author} {\bibfnamefont {F.}~\bibnamefont {Goth}}, \ and\
  \bibinfo {author} {\bibfnamefont {F.~F.}\ \bibnamefont {Assaad}},\ }\href
  {\doibase 10.1103/PhysRevB.97.085144} {\bibfield  {journal} {\bibinfo
  {journal} {Phys. Rev. B}\ }\textbf {\bibinfo {volume} {97}},\ \bibinfo
  {pages} {085144} (\bibinfo {year} {2018})}\BibitemShut {NoStop}%
\bibitem [{\citenamefont {Huang}\ \emph {et~al.}(2024)\citenamefont {Huang},
  \citenamefont {Parthenios}, \citenamefont {Ulybyshev}, \citenamefont {Zhang},
  \citenamefont {Assaad}, \citenamefont {Classen},\ and\ \citenamefont
  {Meng}}]{Cheng24}%
  \BibitemOpen
  \bibfield  {author} {\bibinfo {author} {\bibfnamefont {C.}~\bibnamefont
  {Huang}}, \bibinfo {author} {\bibfnamefont {N.}~\bibnamefont {Parthenios}},
  \bibinfo {author} {\bibfnamefont {M.}~\bibnamefont {Ulybyshev}}, \bibinfo
  {author} {\bibfnamefont {X.}~\bibnamefont {Zhang}}, \bibinfo {author}
  {\bibfnamefont {F.~F.}\ \bibnamefont {Assaad}}, \bibinfo {author}
  {\bibfnamefont {L.}~\bibnamefont {Classen}}, \ and\ \bibinfo {author}
  {\bibfnamefont {Z.~Y.}\ \bibnamefont {Meng}},\ }\href
  {https://arxiv.org/abs/2412.11382} {\bibfield  {journal} {\bibinfo  {journal}
  {axXiv:2412.11382}\ } (\bibinfo {year} {2024})},\ \Eprint
  {http://arxiv.org/abs/2412.11382} {arXiv:2412.11382 [cond-mat.str-el]}
  \BibitemShut {NoStop}%
\bibitem [{\citenamefont {Troyer}\ and\ \citenamefont
  {Wiese}(2005)}]{troyer_computational_2005}%
  \BibitemOpen
  \bibfield  {author} {\bibinfo {author} {\bibfnamefont {M.}~\bibnamefont
  {Troyer}}\ and\ \bibinfo {author} {\bibfnamefont {U.-J.}\ \bibnamefont
  {Wiese}},\ }\href {\doibase 10.1103/PhysRevLett.94.170201} {\bibfield
  {journal} {\bibinfo  {journal} {Physical Review Letters}\ }\textbf {\bibinfo
  {volume} {94}},\ \bibinfo {pages} {170201} (\bibinfo {year} {2005})},\
  \bibinfo {note} {publisher: American Physical Society}\BibitemShut {NoStop}%
\bibitem [{\citenamefont {Weber}\ \emph {et~al.}(2022)\citenamefont {Weber},
  \citenamefont {Honecker}, \citenamefont {Normand}, \citenamefont {Corboz},
  \citenamefont {Mila},\ and\ \citenamefont {Wessel}}]{weber_quantum_2022}%
  \BibitemOpen
  \bibfield  {author} {\bibinfo {author} {\bibfnamefont {L.}~\bibnamefont
  {Weber}}, \bibinfo {author} {\bibfnamefont {A.}~\bibnamefont {Honecker}},
  \bibinfo {author} {\bibfnamefont {B.}~\bibnamefont {Normand}}, \bibinfo
  {author} {\bibfnamefont {P.}~\bibnamefont {Corboz}}, \bibinfo {author}
  {\bibfnamefont {F.}~\bibnamefont {Mila}}, \ and\ \bibinfo {author}
  {\bibfnamefont {S.}~\bibnamefont {Wessel}},\ }\href {\doibase
  10.21468/SciPostPhys.12.2.054} {\bibfield  {journal} {\bibinfo  {journal}
  {SciPost Physics}\ }\textbf {\bibinfo {volume} {12}},\ \bibinfo {pages} {054}
  (\bibinfo {year} {2022})}\BibitemShut {NoStop}%
\bibitem [{\citenamefont {Hangleiter}\ \emph {et~al.}(2020)\citenamefont
  {Hangleiter}, \citenamefont {Roth}, \citenamefont {Nagaj},\ and\
  \citenamefont {Eisert}}]{hangleiter_easing_2020}%
  \BibitemOpen
  \bibfield  {author} {\bibinfo {author} {\bibfnamefont {D.}~\bibnamefont
  {Hangleiter}}, \bibinfo {author} {\bibfnamefont {I.}~\bibnamefont {Roth}},
  \bibinfo {author} {\bibfnamefont {D.}~\bibnamefont {Nagaj}}, \ and\ \bibinfo
  {author} {\bibfnamefont {J.}~\bibnamefont {Eisert}},\ }\href {\doibase
  10.1126/sciadv.abb8341} {\bibfield  {journal} {\bibinfo  {journal} {Science
  Advances}\ }\textbf {\bibinfo {volume} {6}},\ \bibinfo {pages} {eabb8341}
  (\bibinfo {year} {2020})},\ \bibinfo {note} {publisher: American Association
  for the Advancement of Science}\BibitemShut {NoStop}%
\bibitem [{\citenamefont {Alet}\ \emph {et~al.}(2016)\citenamefont {Alet},
  \citenamefont {Damle},\ and\ \citenamefont
  {Pujari}}]{alet_sign-problem-free_2016}%
  \BibitemOpen
  \bibfield  {author} {\bibinfo {author} {\bibfnamefont {F.}~\bibnamefont
  {Alet}}, \bibinfo {author} {\bibfnamefont {K.}~\bibnamefont {Damle}}, \ and\
  \bibinfo {author} {\bibfnamefont {S.}~\bibnamefont {Pujari}},\ }\href
  {\doibase 10.1103/PhysRevLett.117.197203} {\bibfield  {journal} {\bibinfo
  {journal} {Physical Review Letters}\ }\textbf {\bibinfo {volume} {117}},\
  \bibinfo {pages} {197203} (\bibinfo {year} {2016})},\ \bibinfo {note}
  {publisher: American Physical Society}\BibitemShut {NoStop}%
\bibitem [{\citenamefont {Reingruber}\ \emph {et~al.}(2024)\citenamefont
  {Reingruber}, \citenamefont {Caci}, \citenamefont {Wessel},\ and\
  \citenamefont {Richter}}]{reingruber_thermodynamics_2024}%
  \BibitemOpen
  \bibfield  {author} {\bibinfo {author} {\bibfnamefont {A.}~\bibnamefont
  {Reingruber}}, \bibinfo {author} {\bibfnamefont {N.}~\bibnamefont {Caci}},
  \bibinfo {author} {\bibfnamefont {S.}~\bibnamefont {Wessel}}, \ and\ \bibinfo
  {author} {\bibfnamefont {J.}~\bibnamefont {Richter}},\ }\href {\doibase
  10.1103/PhysRevB.109.125120} {\bibfield  {journal} {\bibinfo  {journal}
  {Physical Review B}\ }\textbf {\bibinfo {volume} {109}},\ \bibinfo {pages}
  {125120} (\bibinfo {year} {2024})},\ \bibinfo {note} {publisher: American
  Physical Society}\BibitemShut {NoStop}%
\bibitem [{\citenamefont {Kitaev}(2006)}]{kitaev_anyons_2006}%
  \BibitemOpen
  \bibfield  {author} {\bibinfo {author} {\bibfnamefont {A.}~\bibnamefont
  {Kitaev}},\ }\href {\doibase 10.1016/j.aop.2005.10.005} {\bibfield  {journal}
  {\bibinfo  {journal} {Annals of Physics}\ }\textbf {\bibinfo {volume}
  {321}},\ \bibinfo {pages} {2} (\bibinfo {year} {2006})}\BibitemShut {NoStop}%
\bibitem [{\citenamefont {Giamarchi}(2003)}]{giamarchi_quantum_2003}%
  \BibitemOpen
  \bibfield  {author} {\bibinfo {author} {\bibfnamefont {T.}~\bibnamefont
  {Giamarchi}},\ }\href {\doibase 10.1093/acprof:oso/9780198525004.001.0001}
  {\emph {\bibinfo {title} {Quantum {Physics} in {One} {Dimension}}}}\
  (\bibinfo  {publisher} {Oxford University Press},\ \bibinfo {year}
  {2003})\BibitemShut {NoStop}%
\bibitem [{\citenamefont {Hofmeier}\ \emph {et~al.}(2024)\citenamefont
  {Hofmeier}, \citenamefont {Willsher}, \citenamefont {Seifert},\ and\
  \citenamefont {Knolle}}]{hofmeier_spin-peierls_2024}%
  \BibitemOpen
  \bibfield  {author} {\bibinfo {author} {\bibfnamefont {D.}~\bibnamefont
  {Hofmeier}}, \bibinfo {author} {\bibfnamefont {J.}~\bibnamefont {Willsher}},
  \bibinfo {author} {\bibfnamefont {U.~F.~P.}\ \bibnamefont {Seifert}}, \ and\
  \bibinfo {author} {\bibfnamefont {J.}~\bibnamefont {Knolle}},\ }\href
  {\doibase 10.1103/PhysRevB.110.125130} {\bibfield  {journal} {\bibinfo
  {journal} {Physical Review B}\ }\textbf {\bibinfo {volume} {110}},\ \bibinfo
  {pages} {125130} (\bibinfo {year} {2024})},\ \bibinfo {note} {publisher:
  American Physical Society}\BibitemShut {NoStop}%
\bibitem [{\citenamefont {Saremi}\ and\ \citenamefont
  {Lee}(2007)}]{saremi_quantum_2007}%
  \BibitemOpen
  \bibfield  {author} {\bibinfo {author} {\bibfnamefont {S.}~\bibnamefont
  {Saremi}}\ and\ \bibinfo {author} {\bibfnamefont {P.~A.}\ \bibnamefont
  {Lee}},\ }\href {\doibase 10.1103/PhysRevB.75.165110} {\bibfield  {journal}
  {\bibinfo  {journal} {Physical Review B}\ }\textbf {\bibinfo {volume} {75}},\
  \bibinfo {pages} {165110} (\bibinfo {year} {2007})},\ \bibinfo {note}
  {publisher: American Physical Society}\BibitemShut {NoStop}%
\bibitem [{\citenamefont {Dalla~Piazza}\ \emph {et~al.}(2015)\citenamefont
  {Dalla~Piazza}, \citenamefont {Mourigal}, \citenamefont {Christensen},
  \citenamefont {Nilsen}, \citenamefont {Tregenna-Piggott}, \citenamefont
  {Perring}, \citenamefont {Enderle}, \citenamefont {McMorrow}, \citenamefont
  {Ivanov},\ and\ \citenamefont {Rønnow}}]{dalla_piazza_fractional_2015}%
  \BibitemOpen
  \bibfield  {author} {\bibinfo {author} {\bibfnamefont {B.}~\bibnamefont
  {Dalla~Piazza}}, \bibinfo {author} {\bibfnamefont {M.}~\bibnamefont
  {Mourigal}}, \bibinfo {author} {\bibfnamefont {N.~B.}\ \bibnamefont
  {Christensen}}, \bibinfo {author} {\bibfnamefont {G.~J.}\ \bibnamefont
  {Nilsen}}, \bibinfo {author} {\bibfnamefont {P.}~\bibnamefont
  {Tregenna-Piggott}}, \bibinfo {author} {\bibfnamefont {T.~G.}\ \bibnamefont
  {Perring}}, \bibinfo {author} {\bibfnamefont {M.}~\bibnamefont {Enderle}},
  \bibinfo {author} {\bibfnamefont {D.~F.}\ \bibnamefont {McMorrow}}, \bibinfo
  {author} {\bibfnamefont {D.~A.}\ \bibnamefont {Ivanov}}, \ and\ \bibinfo
  {author} {\bibfnamefont {H.~M.}\ \bibnamefont {Rønnow}},\ }\href {\doibase
  10.1038/nphys3172} {\bibfield  {journal} {\bibinfo  {journal} {Nature
  Physics}\ }\textbf {\bibinfo {volume} {11}},\ \bibinfo {pages} {62} (\bibinfo
  {year} {2015})},\ \bibinfo {note} {publisher: Nature Publishing
  Group}\BibitemShut {NoStop}%
\bibitem [{\citenamefont {Shao}\ \emph {et~al.}(2017)\citenamefont {Shao},
  \citenamefont {Qin}, \citenamefont {Capponi}, \citenamefont {Chesi},
  \citenamefont {Meng},\ and\ \citenamefont {Sandvik}}]{shao_nearly_2017}%
  \BibitemOpen
  \bibfield  {author} {\bibinfo {author} {\bibfnamefont {H.}~\bibnamefont
  {Shao}}, \bibinfo {author} {\bibfnamefont {Y.~Q.}\ \bibnamefont {Qin}},
  \bibinfo {author} {\bibfnamefont {S.}~\bibnamefont {Capponi}}, \bibinfo
  {author} {\bibfnamefont {S.}~\bibnamefont {Chesi}}, \bibinfo {author}
  {\bibfnamefont {Z.~Y.}\ \bibnamefont {Meng}}, \ and\ \bibinfo {author}
  {\bibfnamefont {A.~W.}\ \bibnamefont {Sandvik}},\ }\href {\doibase
  10.1103/PhysRevX.7.041072} {\bibfield  {journal} {\bibinfo  {journal}
  {Physical Review X}\ }\textbf {\bibinfo {volume} {7}},\ \bibinfo {pages}
  {041072} (\bibinfo {year} {2017})},\ \bibinfo {note} {publisher: American
  Physical Society}\BibitemShut {NoStop}%
\bibitem [{\citenamefont {Powalski}\ \emph {et~al.}(2018)\citenamefont
  {Powalski}, \citenamefont {Schmidt},\ and\ \citenamefont
  {Uhrig}}]{powalski_mutually_2018}%
  \BibitemOpen
  \bibfield  {author} {\bibinfo {author} {\bibfnamefont {M.}~\bibnamefont
  {Powalski}}, \bibinfo {author} {\bibfnamefont {K.~P.}\ \bibnamefont
  {Schmidt}}, \ and\ \bibinfo {author} {\bibfnamefont {G.~S.}\ \bibnamefont
  {Uhrig}},\ }\href {\doibase 10.21468/SciPostPhys.4.1.001} {\bibfield
  {journal} {\bibinfo  {journal} {SciPost Physics}\ }\textbf {\bibinfo {volume}
  {4}},\ \bibinfo {pages} {001} (\bibinfo {year} {2018})}\BibitemShut {NoStop}%
\bibitem [{\citenamefont {Sandvik}(2007)}]{sandvik_evidence_2007}%
  \BibitemOpen
  \bibfield  {author} {\bibinfo {author} {\bibfnamefont {A.~W.}\ \bibnamefont
  {Sandvik}},\ }\href {\doibase 10.1103/PhysRevLett.98.227202} {\bibfield
  {journal} {\bibinfo  {journal} {Physical Review Letters}\ }\textbf {\bibinfo
  {volume} {98}},\ \bibinfo {pages} {227202} (\bibinfo {year} {2007})},\
  \bibinfo {note} {publisher: American Physical Society}\BibitemShut {NoStop}%
\bibitem [{\citenamefont {Lieb}(1994)}]{lieb_flux_1994}%
  \BibitemOpen
  \bibfield  {author} {\bibinfo {author} {\bibfnamefont {E.~H.}\ \bibnamefont
  {Lieb}},\ }\href {\doibase 10.1103/PhysRevLett.73.2158} {\bibfield  {journal}
  {\bibinfo  {journal} {Physical Review Letters}\ }\textbf {\bibinfo {volume}
  {73}},\ \bibinfo {pages} {2158} (\bibinfo {year} {1994})},\ \bibinfo {note}
  {publisher: American Physical Society}\BibitemShut {NoStop}%
\bibitem [{\citenamefont {Affleck}\ and\ \citenamefont
  {Marston}(1988)}]{Affleck88}%
  \BibitemOpen
  \bibfield  {author} {\bibinfo {author} {\bibfnamefont {I.}~\bibnamefont
  {Affleck}}\ and\ \bibinfo {author} {\bibfnamefont {J.~B.}\ \bibnamefont
  {Marston}},\ }\href {\doibase 10.1103/PhysRevB.37.3774} {\bibfield  {journal}
  {\bibinfo  {journal} {Phys. Rev. B}\ }\textbf {\bibinfo {volume} {37}},\
  \bibinfo {pages} {3774} (\bibinfo {year} {1988})}\BibitemShut {NoStop}%
\bibitem [{\citenamefont {Lang}\ \emph {et~al.}(2013)\citenamefont {Lang},
  \citenamefont {Meng}, \citenamefont {Muramatsu}, \citenamefont {Wessel},\
  and\ \citenamefont {Assaad}}]{Lang13}%
  \BibitemOpen
  \bibfield  {author} {\bibinfo {author} {\bibfnamefont {T.~C.}\ \bibnamefont
  {Lang}}, \bibinfo {author} {\bibfnamefont {Z.~Y.}\ \bibnamefont {Meng}},
  \bibinfo {author} {\bibfnamefont {A.}~\bibnamefont {Muramatsu}}, \bibinfo
  {author} {\bibfnamefont {S.}~\bibnamefont {Wessel}}, \ and\ \bibinfo {author}
  {\bibfnamefont {F.~F.}\ \bibnamefont {Assaad}},\ }\href {\doibase
  10.1103/PhysRevLett.111.066401} {\bibfield  {journal} {\bibinfo  {journal}
  {Phys. Rev. Lett.}\ }\textbf {\bibinfo {volume} {111}},\ \bibinfo {pages}
  {066401} (\bibinfo {year} {2013})}\BibitemShut {NoStop}%
\bibitem [{\citenamefont {Nasu}\ \emph {et~al.}(2015)\citenamefont {Nasu},
  \citenamefont {Udagawa},\ and\ \citenamefont {Motome}}]{nasu_thermal_2015}%
  \BibitemOpen
  \bibfield  {author} {\bibinfo {author} {\bibfnamefont {J.}~\bibnamefont
  {Nasu}}, \bibinfo {author} {\bibfnamefont {M.}~\bibnamefont {Udagawa}}, \
  and\ \bibinfo {author} {\bibfnamefont {Y.}~\bibnamefont {Motome}},\ }\href
  {\doibase 10.1103/PhysRevB.92.115122} {\bibfield  {journal} {\bibinfo
  {journal} {Physical Review B}\ }\textbf {\bibinfo {volume} {92}},\ \bibinfo
  {pages} {115122} (\bibinfo {year} {2015})},\ \bibinfo {note} {publisher:
  American Physical Society}\BibitemShut {NoStop}%
\bibitem [{\citenamefont {Alexandru}\ \emph {et~al.}(2022)\citenamefont
  {Alexandru}, \citenamefont {Ba\ifmmode~\mbox{\c{s}}\else \c{s}\fi{}ar},
  \citenamefont {Bedaque},\ and\ \citenamefont {Warrington}}]{Alexandru22}%
  \BibitemOpen
  \bibfield  {author} {\bibinfo {author} {\bibfnamefont {A.}~\bibnamefont
  {Alexandru}}, \bibinfo {author} {\bibfnamefont {G.~m.~c.}\ \bibnamefont
  {Ba\ifmmode~\mbox{\c{s}}\else \c{s}\fi{}ar}}, \bibinfo {author}
  {\bibfnamefont {P.~F.}\ \bibnamefont {Bedaque}}, \ and\ \bibinfo {author}
  {\bibfnamefont {N.~C.}\ \bibnamefont {Warrington}},\ }\href {\doibase
  10.1103/RevModPhys.94.015006} {\bibfield  {journal} {\bibinfo  {journal}
  {Rev. Mod. Phys.}\ }\textbf {\bibinfo {volume} {94}},\ \bibinfo {pages}
  {015006} (\bibinfo {year} {2022})}\BibitemShut {NoStop}%
\end{thebibliography}%
\end{document}